\documentclass[useAMS,usenatbib]{mn2e}
\usepackage{psfig}

\input psfig.sty
\usepackage{amsmath}
\usepackage{amssymb}
\usepackage{upgreek}
\usepackage{longtable}
\usepackage[mathscr]{eucal}
\usepackage{graphicx}

\newcommand{\aap}{A\&A}

\newcommand{\aj}{AJ}
\newcommand{\apj}{ApJ}
\newcommand{\apjl}{ApJL}
\newcommand{\apjs}{ApJS}

\newcommand{\araa}{ArA\&A}
\newcommand{\mnras}{MNRAS}

\newcommand{\arcs}{\mbox{\ensuremath{^{\prime\prime}}}}

\title
[Cygnus OB2 was always an Association]
{Constraints on Massive Star Formation: Cygnus OB2 was always an Association}
\author
[Wright et al.]
{Nicholas J. Wright$^1$, Richard J. Parker$^2$, Simon P. Goodwin$^3$ and Jeremy J. Drake$^4$\\
$^{1}$Centre for Astrophysics Research, Science and Technology Research Institute, University of Hertfordshire, Hatfield, AL10 9AB, UK\\
$^{2}$Institute for Astronomy, ETH Z\"urich, Wolfgang-Pauli-Strasse 27, 8093, Z\"urich, Switzerland\\
$^{3}$Department of Physics \& Astronomy, University of Sheffield, Sheffield, S3 7RH, UK\\
$^{4}$Harvard-Smithsonian Center for Astrophysics, 60 Garden Street, Cambridge, MA 02138, USA\\
}

\begin{document}
\maketitle

\begin{abstract}

We examine substructure and mass segregation in the massive OB association Cygnus~OB2 to better understand its initial conditions. Using a well understood {\it Chandra} X-ray selected sample of young stars we find that Cyg~OB2 exhibits considerable physical substructure and has no evidence for mass segregation, both indications that the association is not dynamically evolved. Combined with previous kinematical studies we conclude that Cyg~OB2 is dynamically very young, and what we observe now is very close to its initial conditions: Cyg~OB2 formed as a highly substructured, unbound association with a low volume density ($< 100$ stars~pc$^{-3}$). This is inconsistent with the idea that all stars form in dense, compact clusters. The massive stars in Cyg~OB2 show no evidence for having formed particularly close to one another, nor in regions of higher than average density.  Since Cyg~OB2 contains stars as massive as $\sim$100~M$_\odot$ this result suggests that very massive stars can be born in relatively low-density environments.  This would imply that massive stars in Cyg~OB2 did not form by competitive accretion, or by mergers.

\end{abstract}

\begin{keywords}
stars: formation - kinematics and dynamics - open clusters and associations: individual: Cygnus~OB2
\end{keywords}

\section{Introduction}

The question of whether all stars form in dense clusters is of crucial importance, as it has implications for theories of star formation \citep[e.g.,][]{bonn01}, the processing of binary systems \citep[e.g.,][]{park11b}, and the conditions for the evolution of protoplanetary disks and the formation of planetary systems \citep{armi00,adam06,park12c}. In particular, some theories of massive star formation, such as competitive accretion \citep{bonn01} and stellar mergers, require a dense stellar environment, while other scenarios, such as monolithic collapse \citep[e.g.,][]{york02}, can occur in (and might require) relatively low-density environments \citep[see][for a review]{zinn07}.

There are two competing theories of star formation, and although the reality is likely to be an intermediate combination of the two it can be useful to compare and contrast these theories so that they can be tested. In `clustered star formation' the majority of stars form in dense embedded groups containing thousands to hundreds of thousands of stars within parsec-sized regions \citep[e.g.,][]{lada91b,carp97,krou11}. The feedback-induced expulsion of residual gas left over from the star formation process destroys 90\% of these young clusters within the first 10~Myrs \citep{hill80,lada84,good06}. This widely held view was most prominent advocated by \citet{lada03} and based on the large number of embedded clusters discovered in the near-IR \citep[e.g.,][]{carp00}. However, recent mid-IR observations have challenged this view by revealing that young stellar objects are correlated with the hierarchically structured interstellar medium \citep{gute11} and found over a wide range of stellar surface densities \citep{bres10} suggesting there is no preferred scale of star formation.

What is clear is that only around 10\% of stars find themselves in gas-free bound clusters after a few Myr \citep{lada03}. Many other young stars are found in OB associations: loose, co-moving young stellar groups containing O and/or early B-type stars \citep{blaa64} with a similar stellar content to young star clusters \citep[e.g.,][]{bast10}. Their low stellar mass densities ($<$0.1~M$_\odot$~pc$^{-3}$) imply that they are gravitationally unbound and therefore expanding, which has led to suggestions that they are the expanded remnants of young star clusters disrupted by gas removal \citep{lada91,brow97,krou01}.

Alternatively, in `hierarchical star formation' stars form at a smoothly varying distribution of densities with significant substructure on pc (or greater) scales and denser sub-areas nested within larger, less dense areas \citep[e.g.,][]{scal85,elme06,bast07}. Clusters are formed by merging substructures in the densest subvirial regions \citep{alli09}, whilst low density and unbound regions become OB associations.

These two scenarios provide very different mechanisms for the formation of OB associations, both of which provide clear observational discriminants. In clustered star formation, associations are the expanding remnants of a dynamically evolved dense star cluster. Mixing in the dense star cluster will have erased any initial substructure \citep{scal02,good04,park12b}, but should retain or enhance any mass segregation \citep[which is often observed in bound clusters, e.g.,][]{hill98,stol02}. But in hierarchical star formation associations are dynamically young and should retain any initial substructure \citep{scal02,good04,park12b}, and will only exhibit mass segregation if it was present initially. Thus, substructure (spatial or dynamical) and mass segregation both provide measurable indicators of the level of dynamical evolution within a group of stars, acting as diagnostics of the original physical and dynamical state of the stars when they formed \citep[see][]{park13}. For example \citet{prei99} argued from the kinematics and distribution of stars in the Upper Sco OB association that it must have formed as an association, and very recently, \citet{delg13} used measures of structure and mass segregation to argue for very different dynamical histories for the Berkeley 94 and Berkeley 96 open clusters.

In this paper we attempt to constrain the initial conditions of the formation of the massive OB association Cygnus~OB2 using indicators of dynamical evolution such as substructure and mass segregation. Cyg~OB2 is one of the largest OB associations in our Galaxy with an estimated stellar mass of $\sim$$3 \times 10^4$~M$_\odot$ \citep{drew08,wrig10} and home to many massive stars with masses up to $\sim$100~M$_\odot$ \citep[e.g.,][]{mass91,come02,hans03}, which have an extreme impact on their environment \citep{wrig12}. Furthermore at a distance of only 1.4~kpc \citep{rygl12} it can be studied in sufficient detail to resolve and characterise both high and low-mass stars. This paper is outlined as follows. In Section~2 we introduce the observational sample used for this study and in Section~3 we outline the substructure and mass segregation diagnostics used. In Section~4 we present out results and discuss possible biases, and in Section~5 we discuss our findings in terms of the dynamical and structural evolution of Cyg~OB2 and consider the implications of our results for both Cyg~OB2 and theories of massive star formation.

\section{Sample of young stars in Cyg~OB2}

The observational sample used here is the X-ray selected sample of Cyg~OB2 members presented by \citet{wrig09}. X-ray observations offer a largely unbiased diagnostic of youth that is highly effective in separating young association members from older field stars. This is because pre-main-sequence stars are typically 10--1000 times more luminous in X-rays than main-sequence stars \citep[e.g.,][]{prei05} due to enhanced magnetic activity \citep[for low-mass stars, e.g.,][]{wrig11b} and collisions in strong stellar winds \citep[for high-mass stars, e.g.,][]{naze11}. The only exception to this is A- and late B-type stars that are not believed to emit X-rays \citep[e.g.,][]{schm97}. Another commonly used method for selecting young stars is to use infrared observations to identify stars with circumstellar disks, as recently done by \citet{guar13}. However, in regions such as Cyg~OB2 where the fraction of stars with circumstellar disks is very low \citep[e.g.,][]{alba07,wrig10} and where feedback from the massive O-type stars \citep[e.g.,][]{wrig12} may photoevaporate circumstellar disks and therefore spatially bias the distribution of stars with disks this method could bias studies of the spatial distribution of stars. X-ray observations can however be sensitive to absorption due to neutral hydrogen along the line of sight, the effects of which broadly scale with absorption due to dust, affecting the detection of embedded sources. Fortunately Cyg~OB2 has already dispersed the molecular cloud from which it formed \citep[e.g.,][]{schn06}, with very little evidence for an H~{\sc ii} region in its vicinity \citep{vink08}, and \citet{guar13} noted a dearth of embedded infrared sources toward the centre of the association.

\citet{wrig09} presented a catalogue of X-ray sources in Cyg~OB2 from two observations with the {\it Chandra} X-ray Observatory. The deeper of these two observations was centered on the core of the association and it is the sources from this observation that we use here. \citet{wrig10} studied the properties of these sources, using optical photometry from IPHAS \citep[INT Photometric H$\alpha$ Survey,][]{drew05} to identify and remove foreground contaminants. The masses of stars in the sample range from $\sim$80~M$_\odot$ for Cyg~OB2 \#7, an O3 supergiant, down to 0.1~M$_\odot$. The masses of the high-mass stars were derived from spectroscopy and fitting to evolutionary models \citep{kimi07} and are therefore quite reliable. The masses of individual low-mass stars, while less reliable, are not necessary for the mass segregation diagnostics used here and this is not therefore a concern.

\begin{figure}
\begin{center}
\includegraphics[height=235pt,angle=270]{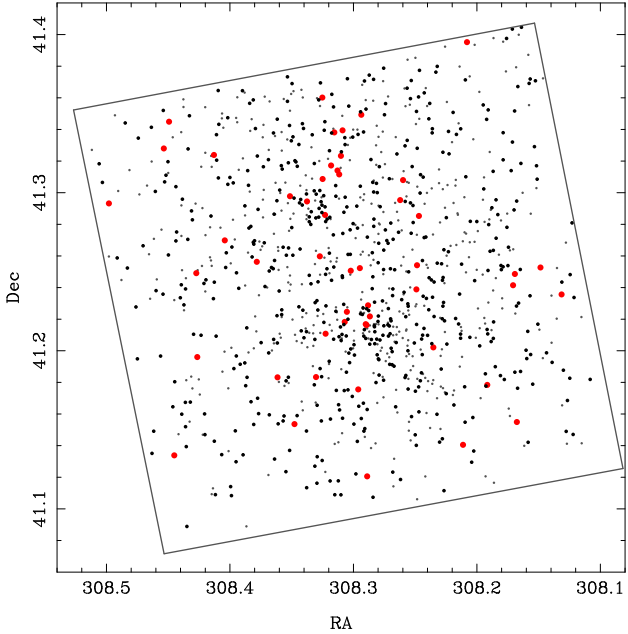}
\caption{Map of the central region of Cyg~OB2 showing the objects in our sample. The 587 stars used for studying substructure and mass segregation are shown as large black dots, with the 50 most massive stars ($M > 11 M_\odot$) shown as red dots. The 445 low-mass stars excluded from this study to avoid spatially-varying incompleteness are shown as small grey dots. The outline of the {\it Chandra} survey area is shown as a grey box.}
\label{map}
\end{center}
\end{figure}

{\it Chandra}'s sensitivity to point sources is highly dependent on the size of the point spread function, which is itself dependent on the distance from the centre of the observation, known as the off-axis angle. This leads to a spatially varying sensitivity that could affect the detection of low-mass stars. Since mass segregation is effectively diagnosing differences in the spatial distribution of stars as a function of their mass it is important that we work with a sample free from mass-dependent spatially varying incompleteness. \citet{wrig10} found that the X-ray luminosity function of our sample was in good agreement with that derived from X-ray studies of other young clusters down to a mass of $\sim$1~M$_\odot$ and that the mass function could be fitted with a slope of $\Gamma = -1.09 \pm 0.13$ (excluding A and B-type stars as described above), in good agreement with the `universal' initial mass function slope of $\Gamma = -1.3 \pm 0.3$ \citep{krou01}. Comparing the distribution of stellar masses with a \citet{krou01} initial mass function we identify the range of masses where the observed mass function deviates from this and which may therefore suffer from spatially varying incompletenesses. We find that the sample is complete in the mass ranges $0.8 \leq M / M_\odot \leq 1.7$ and $M / M_\odot \geq 5$, which we here adopt as our spatially-complete sample for studying mass segregation (hereafter dubbed the `mass function complete' sample). This consists of 587 stars, reduced from the 1032 members of Cyg~OB2 in the full catalog. These stars are distributed over an area of $\sim$0.08~deg$^2$, or $\sim$50~pc$^2$ at the distance of Cyg~OB2. This is equivalent to a surface density of 2--4~stars~arcmin$^{-2}$, significantly below the level at which sample incompleteness effects can bias measures of mass segregation \citep[e.g.,][]{asce09}. Figure~\ref{map} shows the spatial distribution of these sources. Note that the $\sim$50~pc$^2$ surface area shown in Fig~\ref{map} represents around one-third to one-half of the total population of Cyg~OB2.

\section{Methodology}

In this section we outline the substructure and mass segregation diagnostics used in this work, the results of which are presented in Section~\ref{s-results}.

\subsection{The $Q$ parameter measure of cluster structure}

\citet{cart04} pioneered the use of the $Q$-parameter in diagnosing the amount of substructure in star clusters. The $Q$ parameter is defined as $Q = \bar{m} / \bar{s}$, the ratio of the mean edge length of the minimum spanning tree (MST) of all the stars in the cluster, $\bar{m}$, and the mean separation between stars, $\bar{s}$, both normalised as described in \citet{cart04}. Clusters with smooth spatial distribution and central condensation have large $Q$ values, whilst clumpy clusters with significant substructure have small $Q$ values. The advantage of using the $Q$ parameter is that it provides an impartial indication of cluster structure without the need for any arbitrary decisions such as choosing a cluster centre. The normalisation factors also make the parameter independent of the size or density of the star cluster, allowing comparisons between different clusters. While the $Q$ parameter was originally formulated for broadly spherical clusters it can also be adapted to take into account the effects of elongation \citep{bast09}.

\subsection{The $\Lambda_{\rm MSR}$ minimum spanning tree method}

The $\Lambda_{\rm MSR}$ ratio was introduced by \citet{alli09} to provide a quantitative measure of the level of mass segregation with an associated significance \citep[see also][]{olcz11,masc11}. This method uses the length of the MST of a subset of massive stars compared to the mean MST length of many random subsets of low-mass stars. If mass segregation exists in a group of stars then the MST length of the most massive stars will be shorter than the typical MST length of an equal size sample of low-mass stars. \citet{alli09} quantified the mass segregation ratio, $\Lambda_{\rm MSR}$, as

\begin{equation}
\Lambda_{\rm MSR} = \frac{ \langle l_{norm} \rangle}{l_{massive}}
\end{equation}

\noindent
where $l_{massive}$ is the mean MST edge length of $N_{MST}$ massive stars and $\langle l_{norm} \rangle$ is the sample average of the mean MST edge length of $N_{MST}$ stars. The uncertainty on this measure, $\sigma_{norm} / l_{massive}$, can be calculated from Monte Carlo simulations to derive an associated significance. A measurement of $\Lambda_{\rm MSR} \sim 1$ indicates no mass segregation (i.e. the massive stars are distributed in the same way as all other stars), whereas $\Lambda_{\rm MSR} > 1$ indicates mass segregation, with the significance of such a measurement dependent on the uncertainty calculated. This method has particular advantages over other measures of mass segregation based on the radial distributions of the stars in a cluster as it does not rely on defining a cluster centre or any preferred location, a useful feature when studying the spatial distribution of stars in an association that may not have a clear centre.

This method has been well tested on a number of clusters and associations and been shown to produce significant detections of mass segregation in both dynamically evolved clusters and in clusters with known mass segregation \citep[e.g.,][]{alli09,sana10} and also to show a lack of mass segregation in less dynamically evolved groups of stars \citep[e.g.,][]{park11,park12}.

\subsection{The $m - \Sigma$ local stellar surface density method}

An alternative measure of mass segregation based on the local stellar surface density was proposed by \citet{masc11}. If mass segregation exists then the massive stars will be concentrated in denser areas of the cluster and will have higher local surface densities than the general population. This can be seen in a plot of the local surface density, $\Sigma$, versus mass, where $\Sigma = (n-1) / (\pi r_n^2)$, $n$ is the number of stars used to measure the local surface density, and $r_n$ is the distance to the $n^{th}$ nearest neighbour of the star \citep{case85}. We adopt $n=6$ in this work following \citet{masc11} and \citet{case85} who found it to be a good compromise between accurately representing the local density and minimising low-level fluctuations. \citet{masc11} tested this method on the hydrodynamical simulation of star formation by \citet{bonn08}, quantifying the significance of mass segregation using a two-sample Kolmogorov-Smirnov (KS) test of the $\Sigma$ values of the subset compared to the $\Sigma$ values of the entire sample, and found that it provided significant measurements of mass segregation in young clusters. To compare this measurement with that from other clusters we follow \citet{park13} by using the ratio of local surface densities of the 10 most massive stars in the association, $\tilde{\Sigma}_{10}$, to that of all the stars in the association, $\tilde{\Sigma}_{\rm all}$, the local surface density ratio $\Sigma_{\rm LDR} = \tilde{\Sigma}_{\rm 10} / \tilde{\Sigma}_{\rm all}$.

\section{Results}
\label{s-results}

Here we present the results of applying the structural diagnostic $Q$ and both mass segregation diagnostics to our `mass function complete' sample, the implications of which are discussed in Section~\ref{s-discussion}.

\subsection{The Substructure Diagnostic $Q$}
\label{s-q}

We calculate a substructure measure of $Q = 0.34$ for the centre of Cyg~OB2. This is possibly a lower limit due to certain observational effects and the true value is probably 0.4 -- 0.5 (see discussion in Section~\ref{s-biases}). Despite this the true $Q$ value for Cyg~OB2 is still very low. Of the regions examined by \citet{cart04}, only Taurus has such a low $Q$ of 0.47  (although further comparisons between Taurus and Cyg~OB2 should be made cautiously as the two regions are very different and are observed at hugely different distances). Such a low value of $Q$ is almost certainly a signature of a region that is dynamically unevolved as dynamical evolution acts to erase substructure \citep{scal02,good04,park13}.

\subsection{The Mass Segregation Ratio, $\Lambda_{\rm MSR}$}
\label{s-lambda}

The mass segregation ratio $\Lambda_{\rm MSR}$ was calculated for a subset of massive stars of varying size $N_{MST}$ with $\langle l_{norm} \rangle$ calculated from 10,000 random realisations of a random subset of $N_{MST}$ stars drawn from the sample. The distribution of $l_{norm}$ values was then used to calculate $\sigma_{norm}$. This experiment was repeated for multiple values of $N_{MST}$ to identify any possible subset of the massive star population in Cyg~OB2 that might be mass segregated and with different step sizes so that the largest and most significant measurement of mass segregation could be identified.

\begin{figure}
\begin{center}
\includegraphics[height=235pt,angle=270]{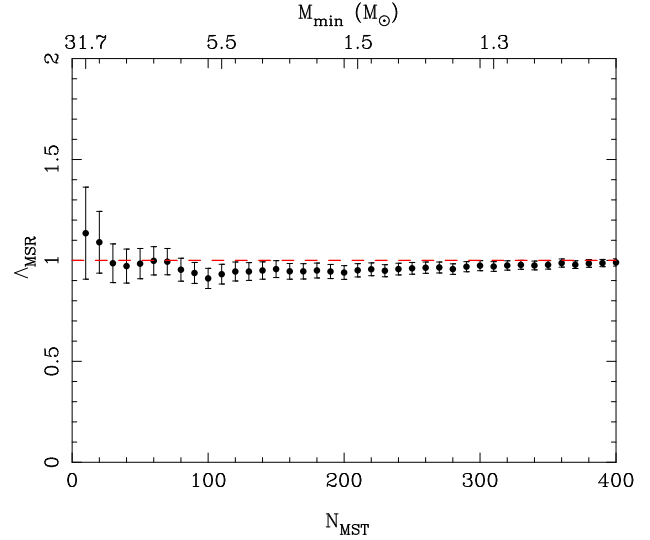}
\caption{Mass segregation ratio, $\Lambda_{\rm MSR}$, for the $N_{MST}$ most massive stars in the centre of Cyg~OB2 in steps of 10 stars using the `mass function complete' sample with 1$\sigma$ error bars. The lowest mass star in each bin is indicated along the top. $\Lambda_{\rm MSR} = 1$, indicating no mass segregation, is shown as a dashed red line.}
\label{results10}
\end{center}
\end{figure}

Figure~\ref{results10} shows the mass segregation ratio $\Lambda_{\rm MSR}$ for the $N_{MST}$ most massive stars in the centre of Cyg~OB2 in steps of 10 stars. The highest mass bin has $\Lambda_{\rm MSR} = 1.14 \pm 0.23$, indicating that the 10 most massive stars (M = 32--80 M$_\odot$) might be slightly more clustered than the average stars in Cyg~OB2, but this result is not significant, deviating from $\Lambda_{\rm MSR} = 1.0$ (no mass segregation) by only 0.6$\sigma$. Increasing $N_{MST}$ produces less significant results and for $N_{MST} > 30$ we find $\Lambda_{\rm MSR} \sim 1$. Adjusting the step value of $N_{MST}$ produces minor changes to the largest value of $\Lambda_{\rm MSR}$, varying from 1.13 to 1.16 as the step size varies from 5-15. However this does not produce more significant results because as $N_{MST}$ increases we lose the ability to pick out structural differences between mass regimes, while if $N_{MST}$ decreases we raise the uncertainty and lower the resulting significance.

This value of $\Lambda_{\rm MSR}$ is significantly lower than that found in other regions \citep[e.g.,][]{alli09,sana10}, both in terms of the absolute measurement and the significance of the measurement. It is also lower than the levels of mass segregation found by \citet{park13} in $N$-body simulations of highly dynamic subvirial clusters (see discussion in Section~\ref{s-discussion}). We therefore conclude that by the $\Lambda_{\rm MSR}$ mass segregation ratio there is no evidence for mass segregation in the centre of Cyg~OB2.

\subsection{The local surface density ratio , $\Sigma_{\rm LDR}$}
\label{s-sigma}

\begin{figure}
\begin{center}
\includegraphics[height=235pt,angle=270]{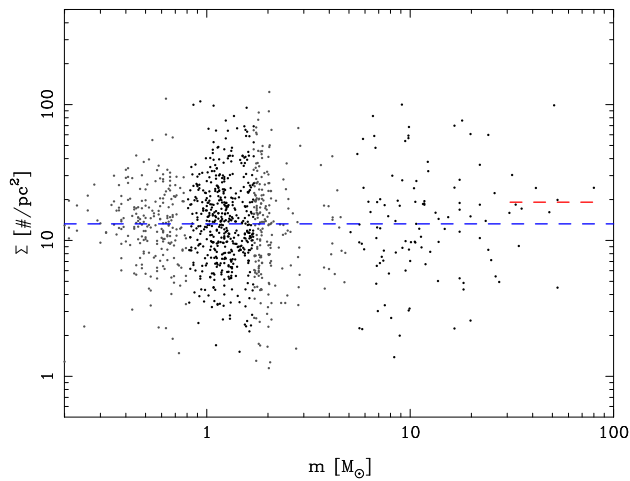}
\caption{The $m-\Sigma$ distribution for all stars in our dataset showing the local surface density for each star plotted against its mass. Our `mass function complete' sample is shown with black dots, while stars excluded from this sample are shown as grey dots. The median surface densities of all the stars in the `mass function complete' sample (blue dashed line) and that of the 10 most massive stars in the sample (red dashed line) are also shown.}
\label{Sigma_results}
\end{center}
\end{figure}

The local surface density, $\Sigma$, for all the stars in our sample is shown in Figure~\ref{Sigma_results}, showing both the full sample and the `mass function complete' subset of the sample.  The spread in $\Sigma$ is approximately two orders of magnitude, lower than the $\sim$3~dex spread measured by \citet{masc11} from their hydrodynamical simulations, but similar to the $\sim$2~dex spread observed by \citet{park12} in $\rho$ Ophiuchi.

The median surface density of the `mass function complete' subset of the sample, $\tilde{\Sigma}_{all} = 13.3$~stars~pc$^{-2}$ is shown, as is the median surface density of the 10 most massive stars in the sample $\tilde{\Sigma}_{10} = 19.1$~stars~pc$^{-2}$. This difference is not significant however, with a two dimensional KS test returning a p-value of 0.24 that the two subsets share the same parent distribution. The local surface density ratio for Cyg~OB2 is then $\Sigma_{\rm LDR} = 1.44$, much lower than the values of $\Sigma_{\rm LDR}$ found by \citet{park13} in their $N$-body simulations of both subvirial (bound) and supervirial (unbound) dense clusters. Given the large number of massive stars in Cyg~OB2 it might be considered restrictive to only use the 10 most massive stars for this diagnostic, though there is a fine balance between sensitivity to the most massive stars and the statistical significance of the result afforded by the sample size. Recalculating the local surface density ratio using the 20 (30) most massive stars changes the ratio to $\Sigma_{\rm LDR} = 1.34$ (1.28), a very small change which does not alter the overall result. We conclude that the massive stars in the centre of Cyg~OB2 are not in regions of significantly higher local density than the low-mass stars, and are therefore not mass segregated according to this ratio.

\subsection{Possible biases}
\label{s-biases}

Our observations suggest that there is no significant evidence for mass segregation in Cyg~OB2 and that the association exhibits considerable substructure. These results are based on the spatial distribution of stars, both that of the entire sample and that of the IMF complete sample. Anything that could affect our ability to detect and characterise stars at different spatial densities or stars of different masses could therefore bias these results. We consider such possible biases here and attempt to assess their impact on our results.

One possible bias is evident from the positions of stars in Figure~\ref{map}, which reveals a cross-shape of low stellar density due to the gap between {\it Chandra}'s CCDs. This chip gap of 11\arcs\ is partly smoothed out by the Lissajous dither pattern used by the observatory, but will leave an area of low sensitivity between CCDs. While this will not affect the positions of the OB stars (which are known from other observations) and therefore the level of mass segregation, it may induce structural features that will artificially decrease $Q$. To test the importance of this effect we simulated fractal datasets with and without a cross in the centre of the image. For ten different realisations of a region with 1000 stars in a 3D fractal with a fractal dimension of 2.0 we find that the `true' 2D value of $Q$ varies between 0.42 and 0.63 (typically $\sim$0.5). Placing a `cross' with a size of 10 per cent of the total size of the region (a conservative over-estimation) typically lowers the measured 2D $Q$ value by around 0.1 -- giving a range of $Q$ between 0.27 and 0.60 (note that in one case the $Q$-value increases by only 0.06). Therefore the measured $Q=0.34$ for Cyg~OB2 is likely underestimated slightly and the true $Q$-value is probably 0.4 -- 0.5, still very low.

Another bias that could affect our sample is contamination of the sample by non-members of Cyg~OB2. These objects would be randomly distributed across the field and would appear as low-mass stars (since all the high-mass stars in Cyg~OB2 have been spectroscopically identified). {\em Significant} contamination would effect the values of all of our quantitative measures. The effect of adding randomly positioned contaminants is to smooth out density differences, effectively pushing $Q$ towards 0.8 (i.e. smoothly distributed) and pushing $\Sigma_{\rm LDR}$ towards unity (i.e. to preferentially increase the densities of low-surface density regions). The potential effects of contamination on $\Lambda_{\rm MSR}$ are subtle, and it could artificially increase or decrease $\Lambda_{\rm MSR}$ depending on what the true underlying distribution is. However, the very low measured value of $Q$ shows that no significant randomly distributed component is present (otherwise $Q$ would not be so low).  Therefore we conclude that contamination is not significant in this sample.

Finally we note that the effects of variable extinction are unlikely to have a significant effect on our results. \citet{bast09} studied how incompletenesses due to extinction can affect the resulting $Q$ parameter, causing the measured value to be lower by 0.04--0.08 if 20--50\% of the sources are undetected due to variable extinction. This result was supported by a similar study by \citet{park12b} who also found the same was true when calculating $\Sigma$, i.e., only when an unphysically larger number of stars are undetected due to extinction do such structural diagnostics become unreliable. It is worth reiterating that we do not expect a significant loss of sources due to variable extinction since \citet{guar13} did not detect many embedded sources in Cyg~OB2 from their deep infrared study.

\section{Discussion}
\label{s-discussion}

Cyg~OB2 is an association with a total mass estimated to be $3 \times 10^4$~M$_\odot$ \citep{drew08,wrig10} spread over an area of at least 50~pc$^2$  and surrounded by (but not embedded within) the molecular cloud complex Cygnus~X with a gas mass of $3 \times 10^6$~M$_\odot$ \citep[adjusted for a distance of 1.4~kpc,][]{schn06}. Based on the results from this paper we can make several statements:\\

\noindent 1. The centre of Cyg~OB2 shows a significant degree of substructure with a true 2D $Q$-value of 0.4--0.5 (see Section~\ref{s-q}).\\
2. Cyg~OB2 shows no evidence that the massive stars are distributed any differently to the low-mass stars (as measured by $\Lambda_{\rm MSR}$, see Section~\ref{s-lambda}).\\
3. Cyg~OB2 shows no evidence that the massive stars are in regions of higher local density than the low-mass stars (as measured by $\Sigma_{\rm LDR}$, see Section~\ref{s-sigma}).\\

\noindent Putting together all of this evidence we argue that {\em Cyg~OB2 has always been a substructured, unbound association}.

The significant degree of spatial substructure as measured by $Q$ strongly suggests that Cyg~OB2 is dynamically young. That is, it has not been able to mix in phase space and retains the imprint of its initial conditions \citep[a picture supported by evidence of physical and dynamical substructure in Cyg~OB2, e.g.,][]{wrig12b,guar13}. Previous studies have found that substructure is only ever erased \citep{scal02,good04}. In particular, \citet{park13} find that $Q$ tends to stay the same or increase in the vast majority of simulations, although in some initially smooth and unbound regions substructure can increase very slightly to $\sim$0.8 and then quickly falls to $\sim$0.6 before remaining roughly constant. This is due to sub-regions with locally similar velocities being able to `condense' from an initially smooth distribution. The decrease in $Q$ is however small and we also believe such smooth initial conditions to be highly unphysical. Therefore, the current value of $Q$ is an upper limit on the initial value of $Q$. The fact that we see a low current value of $Q$ means that {\em Cyg~OB2 has always contained significant substructure}.

The lack of any evidence for mass segregation is extremely interesting. That $\Lambda_{\rm MSR} \sim 1$ shows that the massive stars are not closer together than would be expected from a random selection of low-mass stars. \citet{park13} find that in bound `clusters' $\Lambda_{\rm MSR}$ tends to increase (though it can go down due to the dynamical decay of higher-order Trapezium-like systems), but in unbound regions $\Lambda_{\rm MSR}$ retains its initial value (as the massive stars have no chance to group together). The velocity dispersion of Cyg~OB2 suggests the region is gravitationally unbound \citep[see][and erratum]{kimi07} and therefore that $\Lambda_{\rm MSR}$ was always unity -- i.e. {\em the massive stars in Cyg~OB2 were never grouped together more closely.}

The local surface density around the massive stars as measured by $\Sigma_{\rm LDR}$ is also statistically the same as that around low-mass stars. \citet{park13} show that in bound \emph{and} unbound regions $\Sigma_{\rm LDR}$ always tends to increase.  This is because the massive stars act as a local potential well into which they can attract a retinue of low-mass stars increasing their local surface density. Therefore $\Sigma_{\rm LDR}$ is a lower limit on the initial $\Sigma_{\rm LDR}$ which increases with dynamical age.  This again suggests that Cyg~OB2 is dynamically young as the massive stars have had no (dynamical) time to attract a local retinue (alternatively they have had time, but Cyg~OB2 started with the massive stars in significantly less locally dense regions), i.e. {\em the massive stars in Cyg~OB2 did not form in locally overdense regions.}

In particular, given the age of around 3--5~Myr of Cyg~OB2 \citep{wrig10}, and comparing with the simulations of \citet{park13} we find that only unbound (supervirial) regions with initial volume densities of $< 100$ stars~pc$^{-3}$ are of low enough density for the massive stars to fail to gather a retinue in a few~Myr.  In collapsing, or in higher surface density regions (assuming the third dimension is roughly the same as the observed two dimensions) $\Sigma_{\rm LDR}$ is always found to increase significantly in a few~Myr. The surface density of the observed region is several hundred stars~pc$^{-2}$ (extrapolating to a full IMF), and if the third dimension is roughly the same as the two observed dimensions this suggests an average volume density in this region of around 100 stars~pc$^{-3}$ -- in good agreement with the theoretical argument. All the evidence above suggests that Cyg~OB2 is dynamically young which would be expected if it was {\em born} unbound.  

\subsection{Implications for theories of massive star formation}

Cyg~OB2 contains a number of very massive stars with masses of $\sim$100~M$_\odot$ \citep[e.g.,][]{mass91,kimi07}, particularly the blue hypergiant Cyg~OB2 \#12, which is reported to have a mass of 110~M$_\odot$ \citep{clar12}. The presence of such massive stars is consistent with estimates of the total stellar mass of Cyg~OB2 of $\sim$$3 \times 10^4$~M$_\odot$ and make it comparable with some of the most massive star clusters in our Galaxy such as NGC~3603 or Westerlund~1. Therefore the conditions under which Cyg~OB2 and its massive stars formed is particularly important for our understanding of how such stars form and acts as a constraint for theories of massive star formation.

There are a number of theories for how massive stars form and build up their considerable masses, ranging from scaled up versions of low-mass star formation \citep[e.g.,][]{shu87,mcke03}, collisions or mergers in the cores of dense clusters \citep{zinn07} and relatively dynamic theories where environment plays a significant role \citep[e.g.,][]{bonn04}. The concept of {\it competitive accretion} is a particular example of the latter theory and suggests that high-mass stars begin their lives as relatively low-mass molecular cores but are able to accrete considerably more matter than other stars due to their preferential positions in the centres of dense clusters where the gravitational potentials are highest \citep[e.g.,][]{zinn82,lars92,bonn04}. This requires that massive stars are only born in dense massive clusters, and should also be preferentially found in the centres of these clusters, i.e. clusters should exhibit a level of {\it primordial mass segregation} that cannot be explained by dynamical means \citep{bonn98}.

Our results suggest that the massive stars in Cyg~OB2 did not form close together (either in a single cluster, or in a few clusters as this would be retained in $\Lambda_{\rm MSR}$), nor did they form in locally overdense regions (which would be indicated by a high $\Sigma_{\rm LDR}$).  The presence of stars as massive as 100~M$_\odot$ in Cyg~OB2 is inconsistent with the idea that massive stars can only form in dense clusters. This argues against theories that require massive stars to only form in dense massive clusters, such as the theory of competitive accretion \citep[e.g.,][]{bonn04} or the formation of massive stars by mergers \citep{zinn07}, as the only mechanisms by which massive stars form.

\subsection{Implications for our understanding of Cyg~OB2}

We suggest it is highly unlikely that Cyg~OB2 was ever a single compact cluster which is in the process of destroying itself post gas-expulsion.  In such a case we would not expect to see spatial substructure, and we might expect to see some evidence of the (primordial or dynamical) mass segregation of the initial cluster retained. By far the best explanation for the observed properties of Cyg~OB2 is that we are seeing the region now very much as it formed, as an unbound association with a relatively low surface density.

Such an interpretation of the initial conditions provides a natural explanation for the large range of stellar ages measured in Cyg~OB2.  This was first hinted at by \citet{mass91} who noted the presence of evolved supergiants alongside the high-mass main sequence population in Cyg~OB2, and this has since been confirmed by other authors \citep[e.g.,][]{hans03,come12}. Furthermore amongst the lower-mass population \citet{drew08} uncovered a 5--7~Myr old population of A-type stars and \citet{wrig10} found a spread of ages of 3--5~Myr. Whilst there is considerable debate about the reality of age spreads amongst low-mass stars \citep[e.g.][]{pall99,jeff11}, the existence of multiple age populations inferred from OB stars are less prone to such uncertainties, and the evidence from different mass ranges supports the view that Cyg~OB2 is not a simple coeval population.

Our finding that Cyg~OB2 was born in a highly substructured and low density arrangement suggests that the stars were most likely born over a much larger area, $>$10~pc, than the typical compact size of young star clusters, $\sim$1--2~pc. The observed range of stellar ages could therefore be considered as due to a series of discrete and hierarchical star formation events that have since expanded and overlapped. Indeed, it would seem unlikely to not have age spreads of a few Myr over a region around 10~pc across.

\subsection{What is the true 3D structure of Cyg~OB2?}

As is almost always true in astronomy, our observations of Cyg~OB2 are a 2D projection of a 3D region.  When dealing with spherical and gravitationally bound `clusters', the assumption that the third dimension is very similar to the two observed dimensions is probably very reasonable. However, the observations of Cyg~OB2 show significant substructure (a very low-$Q$), and combined with the high (unbound) velocity dispersion and significant age spreads suggest a poorly-mixed, dynamically young region.  This raises the question of the possible importance of the true 3D shape of Cyg~OB2 and projection effects.  It is extremely difficult to imagine how projection effects could give either a low $Q$ value or a low $\Sigma_{\rm LDR}$ value if they were not the true values (its effects on $\Lambda_{\rm MSR}$ are not obvious), but the degree to which it could alter various structure parameters is unclear.  We will examine this in more detail in a future paper.

\section{Conclusions}

The question of whether all stars form in dense clusters has fundamental ramifications for theories of star formation, the formation mechanisms of high-mass stars and whether clusters represent a fundamental unit of star formation. In this paper we have studied the structure of the massive Cyg~OB2 association in an attempt to constrain its initial conditions.

To determine the amount of dynamical evolution we have studied the level of physical substructure and searched for evidence of mass segregation in Cyg~OB2 using a well-characterised X-ray selected sample of young stars down to 1~M$_\odot$. We used the $Q$ parameter to diagnose substructure \citep{cart04} and two independent measures of mass segregation, $\Lambda_{\rm MSR}$ \citep{alli09} and $\Sigma_{\rm LDR}$ \citep{masc11,park13}.  Our results show that Cyg~OB2 has considerable substructure and is not mass segregated, both indications that the association is dynamically young \citep[see][]{park13}. We therefore infer that the initial conditions of Cyg~OB2 were: \\
1) Cyg~OB2 formed as a relatively low-density, highly substructured, globally unbound association and has changed little in its bulk properties since its formation.\\
2) The massive stars in Cyg~OB2 did not form close together, nor did they form in regions of higher than average local surface / volume density.

The overall conclusion is that Cyg~OB2 formed very much as we see it today and was not born as a dense cluster. Since Cyg~OB2 contains many very massive stars, including at least two stars as massive as $\sim$100~M$_\odot$, this allows us to constrain the sites and conditions under which massive stars form. The formation of these massive stars in a low density environment is inconsistent with the idea that massive stars are only born in dense clusters where the deep potential well caused by a massive and dense star cluster allows the massive stars to attract and accrete sufficient mass to reach such high stellar masses.  It is also extremely difficult to imagine any environment in the young Cyg~OB2 that would allow mergers to occur.  Any theory of massive star formation must therefore be able to explain how stars as massive as $\sim$100~M$_\odot$ can form in a low density association such as Cyg~OB2.

The total mass and content of massive stars make Cyg~OB2 comparable to some of the most massive star clusters in our Galaxy, such as NGC~3603 or Westerlund~1, yet as an association its members are now, and we argue always have been, spread over a much larger area. The question of whether two such similar populations of stars as Cyg~OB2 and Westerlund~1 (both with similar total masses and initial mass functions) formed in such different spatial configurations as they appear now, or whether they formed in the same manner and have since evolved in different directions, is an important issue for theories of star formation.

This study was enabled by the high spatial resolution of {\it Chandra} X-ray observations, which provide an unbiased and quasi-complete sample of low mass stars in Cyg~OB2. The larger {\it Chandra} Legacy Survey of Cyg~OB2 will allow this study to be extended over a much larger area in the future and with a larger number of stars. Kinematical observations such as radial velocities and proper motions from upcoming facilities such as {\it Gaia} and associated ground-based spectroscopic surveys can be used to test our results by searching for and quantifying the level of energy equipartition and dynamical substructure. There is also considerable potential for combining kinematical observations with spatial diagnostics such as those explored in this paper, which we plan to address in a future paper.

\section{Acknowledgments}

The authors would like to thank Janet Drew, Geert Barentsen, and Mario Guarcello for stimulating discussions and helpful comments on this paper. We also thank the anonymous referee for constructive comments that helped improve the paper. NJW acknowledges a Royal Astronomical Society Research Fellowship. This work is based on ideas and discussions as part of an International Team at the International Space Science Institute in Bern, Switzerland.

\bibliographystyle{mn2e}

\end{document}